\documentclass[
a4paper, 
10pt,    
twoside, 
]{LTJournalArticle}

\usepackage{amsmath}
\usepackage{amsfonts}
\usepackage{amssymb}
\usepackage{amstext}
\usepackage{amsthm}

\newtheorem{remark}{Remark}
\newcounter{RomanNumber}

\usepackage{algorithm}
\usepackage{algorithmic}
\usepackage{color}
\usepackage{makecell}
\usepackage{multicol}

\title{The AECM Algorithm for Deterministic Maximum Likelihood Direction Finding in the Presence of Gaussian Mixture Noise}

\author{Mingyan Gong and Bin Lyu
}
\date{}

\renewcommand{\maketitlehookd}{
	\begin{abstract}
		\noindent Gaussian mixture noise can model non-Gaussian noise and also be used when outliers are present in measurements. For deterministic maximum likelihood direction finding in Gaussian mixture noise, the Space-Alternating Generalized Expectation-maximization (SAGE) algorithm, an extension of the expectation-maximization algorithm, was applied and designed by Kozick and Sadler twenty odd years ago, which simultaneously updates DOA estimates at each iteration and cannot properly converge under unequal signal powers. In this paper, the Alternating Expectation-Conditional Maximization (AECM) algorithm, an extension of the SAGE algorithm, is applied and designed, which utilizes multiple less informative versions of the complete data and the golden section search method to update DOA estimates at each iteration sequentially (one by one). Theoretical analysis reveals that the AECM algorithm possesses nearly the same computational complexity of each iteration as the SAGE algorithm. However, simulation results prove that the AECM algorithm exhibits faster stable convergence and is more attractive in computation. Importantly, the AECM algorithm, unlike existing subspace based robust DOA estimation methods, can still be used when signals are coherent. 

        \par\textbf{Keywords: }Array signal processing, DOA estimation, EM algorithm, Gaussian mixture model, impulsive noise, radar signal processing.
	\end{abstract}
}

\begin{document}

\maketitle 

\section{Introduction}

Direction of arrival (DOA) estimation is a crucial research topic in array signal processing and is relevant to diverse applications, such as radar, sonar, and wireless communications \cite{Godara}. In DOA estimation, Gaussian noise is the most common noise model \cite{Miller,Krim}, but the Gaussian distribution may be unsuitable for a number of practical noises, e.g., impulsive noise. To this end, several non-Gaussian distributions have been developed. For example, the symmetric $\alpha$-stable ($S\alpha S$) process has been adopted to model impulsive noise or measurements containing outliers \cite{Shao,Samorodnitsky,Swami,Nikias}, and many robust methods have been proposed, including ROC-MUSIC \cite{Tsakalides}, FLOM-MUSIC \cite{THLiu}, methods based on spatial sign and rank \cite{Visuri}, $\ell_p$-MUSIC \cite{Zeng}, and some recent methods \cite{Raj,LiaoYP,Song}. In particular, Gaussian mixture noise attracts a lot of attention due to its wide applicability.

Of course, Gaussian mixture noise is based on well-known Gaussian mixture models (GMMs) that are used to model the distributions of a wide variety of random phenomena and have been applied to the areas of applied statistics, e.g., pattern recognition, machine learning, and signal processing \cite{GJMcLachlan}. GMMs involve the mixture density estimation problem, which causes a number of techniques. Compared to other techniques, the technique of maximum likelihood (ML) attracts more attention with increasingly powerful computers and increasingly sophisticated numerical methods \cite{Redner}. In order to efficiently obtain ML based parameter estimates, it appears that the most prominent numerical method is the expectation-maximization (EM) algorithm \cite{Dempster, GMcLachlan}.

Instead of directly manipulating the objective or actual log-likelihood function (LLF) of measurements (known as incomplete data), the EM algorithm constructs flexible complete data, which contains missing data, and utilizes the complete-data LLF to update parameter estimates at each iteration. In such a way, the actual LLF can still be increased. In general, the complete-data LLF is more tractable than the actual (or incomplete-data) LLF, which makes the EM algorithm attractive in computation. As the name suggests, its each iteration is composed of one expectation step (E-step) as well as one maximization step (M-step), namely, one EM-pair. However, the EM algorithm simultaneously updates all parameter estimates by maximizing the expected complete-data LLF with respect to all parameters, which results in two drawbacks: 1) very slow convergence and 2) complicated maximization in complex applications. To mitigate both, the Space-Alternating Generalized EM (SAGE) algorithm has been proposed \cite{Fessler}, which utilizes multiple versions of the complete data and each is only associated with updating partial parameter estimates and one EM-pair. In other words, the SAGE algorithm utilizes multiple EM-pairs to sequentially update parameter estimates at each iteration. 

Because of the fast convergence and low complexity, Kozick and Sadler have applied the SAGE algorithm for solving the deterministic ML direction finding problem in Gaussian mixture noise \cite{Kozick}. To our best knowledge, this SAGE algorithm is the only efficient method for this complex optimization problem and therefore has been cited by nearly all articles about DOA estimation in impulsive noise. However, the SAGE algorithm employs the more informative complete data to simultaneously update DOA estimates at each iteration, which cannot yield the fastest convergence. More importantly, the SAGE algorithm requires maximizing the expected complete-data LLF with respect to partial parameters at each M-step and cannot properly converge under unequal signal powers. Specifically, all or both of its sequences of different DOA estimates get close to the true DOA of the source with the largest or larger signal power.

In this paper, in order to address the issue of improper convergence from the SAGE algorithm, the Alternating Expectation-Conditional Maximization (AECM) algorithm \cite{Meng}, an extension of the SAGE algorithm, is applied and designed. To be precise, the AECM algorithm utilizes multiple less informative versions of the complete data (or EM-cycles) and the golden section search method to update DOA estimates at each iteration sequentially (one by one). Furthermore, it utilizes conditional M-steps (CM-steps) to reduce the estimate differences of each DOA between adjacent iterations. Here, M-step and EM-pair in SAGE are extended to CM-step and EM-cycle in AECM, respectively, so the AECM algorithm does not need to maximize the expected complete-data LLF with respect to partial parameters at each CM-step. Theoretical analysis reveals that the AECM algorithm possesses nearly the same computational complexity of each iteration as the SAGE algorithm. However, simulation results prove that the AECM algorithm exhibits faster stable convergence and is more attractive in computation. Importantly, the AECM algorithm, unlike existing subspace based robust DOA estimation methods, can still be used when signals are coherent. 

\emph{Notations}: $(\cdot)^T$, $(\cdot)^H$, $\vert\cdot\vert$, and $\mathrm{Tr}\{\cdot\}$ mean the transpose, conjugate transpose, modulus, and trace operators, respectively. $\jmath$ denotes the imaginary unit and $[\textbf{a}]_n$ stands for the $n$th element of vector $\mathbf{a}$. $\boldsymbol{\Sigma}\succ\mathbf{0}_N$ represents that the conjugate symmetric matrix $\boldsymbol{\Sigma}$ of order $N$ is positive definite while $\boldsymbol{\Sigma}^{-1}$ is the inverse matrix of $\boldsymbol{\Sigma}$. $\mathcal{E}\{\cdot\}$ and $\mathcal{D}\{\cdot\}$ denote the expectation and covariance operators, respectively. $\hat{\theta}$ is the estimate of unknown parameter $\theta$, $\bar{\theta}$ is the $\hat{\theta}$ updated at the previous iteration, and $\breve{\theta}$ is the $\hat{\theta}$ updated at the current iteration. $\mathrm{diag}\{a_1,\dots,a_N\}$ denotes the $N\times N$ diagonal matrix with diagonal elements $a_1,\dots,a_N$. Moreover, $\mathbf{1}=[1~\cdots~1]^T$ and $\mathbf{0}=[0~\cdots~0]^T$.

\section{Deterministic ML Direction Finding Problem}

Let us assume a uniformly spaced linear array of $N$ omnidirectional sensors receiving $M~(M<N)$ narrowband plane waves impinging with distinct directions $\theta_m$'s, where $\theta_m\in(0,\pi)$ is the DOA of the $m$th source wave. For simplicity, let the spacing between adjacent sensors be the half-wavelength of these source waves. Hence, the sensor array output signal can be modeled as
\begin{equation} \label{1}  
\mathbf{y}(t)=\sum_{m=1}^M\mathbf{a}(\theta_m)s_m(t)+\mathbf{v}(t)=\mathbf{A}(\boldsymbol{\theta})\mathbf{s}(t)+\mathbf{v}(t),
\end{equation}
where $\mathbf{a}(\theta_m)=[1~e^{-\jmath\pi\cos(\theta_m)}~\cdots~e^{-\jmath(N-1)\pi\cos(\theta_m)}]^T\in\mathbb{C}^N$ denotes the steering vector of the $m$th source wave, $\boldsymbol{\theta}=[\theta_1~\cdots~\theta_M]^T$, and $\mathbf{A}(\boldsymbol{\theta})=[\mathbf{a}(\theta_1)~\cdots~\mathbf{a}(\theta_M)]$ stands for the array steering matrix. Moreover, $\mathbf{s}(t)=[s_1(t)~\cdots~s_M(t)]^T\in\mathbb{C}^M$ is the source waveform vector.

\subsection{Gaussian Mixture Noise}

In \eqref{1}, $\mathbf{v}(t)$ represents a Gaussian mixture noise vector with mean zero and the probability dense function (PDF) of the $n$th element $v_n(t)$ in $\mathbf{v}(t)$ is
\begin{equation} \label{2}  
p\big(v_n(t);\boldsymbol{\lambda},\boldsymbol{\sigma}\big)
=\sum_{l=1}^L\frac{\lambda_l}{\pi\sigma^2_l}\exp\Big[-\frac{\vert v_n(t)\vert^2}{\sigma^2_l}\Big],
\end{equation}
where $L$ denotes the number of mixture components, $\lambda_l$ and $\sigma^2_l$ are the mixing proportion and variance of the $l$th component, respectively. Moreover, $\boldsymbol{\lambda}=[\lambda_1~\cdots~ \lambda_L]^T>\mathbf{0}$, $\boldsymbol{\lambda}^T\mathbf{1}=1$, and $\boldsymbol{\sigma}=[\sigma_1~\cdots~\sigma_L]^T>\mathbf{0}$. For analytical simplicity, let $L$ be fixed and known \cite{Kozick}.

We adopt the deterministic source signal model \cite{Ziskind}, so $\mathbf{s}(t)$ in \eqref{1} is \textbf{deterministic but unknown}, and the PDF of the $n$th element $y_n(t)$ in $\mathbf{y}(t)$ is
\begin{equation} \label{3}  
p\big(y_n(t);\boldsymbol{\Phi}(t)\big)=
\sum_{l=1}^L\frac{\lambda_l}{\pi\sigma^2_l}\exp\Big[\frac{\big\vert y_n(t)-[\mathbf{A}(\boldsymbol{\theta})\mathbf{s}(t)]_n\big\vert^2}{-\sigma^2_l}\Big],
\end{equation} 
where $\boldsymbol{\Phi}(t)=(\boldsymbol{\theta},\mathbf{s}(t),\boldsymbol{\lambda},\boldsymbol{\sigma}\big)$. Since the $v_n(t)$'s received at different sensors are assumed to be mutually independent, the PDF of $\mathbf{y}(t)$ is
\begin{equation} \label{4} 
p\big(\mathbf{y}(t);\boldsymbol{\Phi}(t)\big)=\prod_{n=1}^Np\big(y_n(t);\boldsymbol{\Phi}(t)\big).
\end{equation}

\subsection{Problem Description}

Practically speaking, we need to sample $\mathbf{y}(t)$ and collect statistically independent measurements or snapshots:
\begin{equation}  \label{5}  
\mathbf{y}(t)=\mathbf{A}\mathbf{s}(t)+\mathbf{v}(t),t=1,\dots,T
\end{equation}
where $T$ represents the number of measurements or snapshots, $P_m=(1/T)\sum_{t=1}^T\vert s_m(t)\vert^2$ stands for the signal power of the $m$th source, and $\mathbf{A}(\boldsymbol{\theta})$ is replaced with $\mathbf{A}$ for notational simplicity. After that, we get hold of the DOA estimates $\hat{\theta}_m$'s by manipulating the measurements or snapshots $\mathbf{Y}$ with $\mathbf{Y}=[\mathbf{y}(1)~\cdots~\mathbf{y}(T)]$. Toward this goal, we adopt the technique of ML \cite{Kay} and write the LLF of $\mathbf{Y}$ as
\begin{equation}  \label{6}  
f(\boldsymbol{\Phi})=\ln p(\mathbf{Y};\boldsymbol{\Phi})=\sum_{t=1}^T\sum_{n=1}^N\ln p\big(y_n(t);\boldsymbol{\Phi}(t)\big),
\end{equation}
where $\boldsymbol{\Phi}=(\boldsymbol{\theta},\mathbf{S},\boldsymbol{\lambda},\boldsymbol{\sigma})$ with $\mathbf{S}=[\mathbf{s}(1)~\cdots~\mathbf{s}(T)]$. Awkwardly, we must solve the deterministic ML direction finding problem:
\begin{equation}   \label{7}    
\max_{\boldsymbol{\theta}\in(0,\pi)^M,\mathbf{S},\boldsymbol{\lambda}>\mathbf{0},\boldsymbol{\lambda}^T\mathbf{1}=1,\boldsymbol{\sigma}>\mathbf{0}}f(\boldsymbol{\Phi}).
\end{equation}

In order for solving problem \eqref{7} efficiently, the SAGE algorithm was applied and designed by Kozick and Sadler twenty odd years ago \cite{Kozick}, but its convergence can still be improved and it cannot properly converge under unequal signal powers, i.e., $P_1=\cdots=P_M$ is not satisfied. In the next section, we introduce this SAGE algorithm along with a clear derivation.

\section{SAGE Algorithm}

It is well-known that the EM algorithm is very suitable for efficiently estimating the ML parameters of a GMM \cite{Redner}. Thus, Kozick and Sadler designed the SAGE algorithm, an extension of the EM algorithm, to solve problem \eqref{7}.

\subsection{Complete Data}

EM-type algorithms require complete data, which contains missing data \cite{GMcLachlan}. According to the ML fitting of GMMs, we introduce the missing data or discrete random variable $h_n(t)\in\{1,\dots,L\}$, which can be seen as the component-label of $v_n(t)$, and rewrite Eq. \eqref{2} as
\begin{eqnarray}  \label{8}  
p\big(v_n(t);\boldsymbol{\lambda},\boldsymbol{\sigma}\big)&=&\sum_{h_n(t)=1}^L\Big[p\big(h_n(t);\lambda_{h_n(t)}\big)\times\nonumber\\
&&\underbrace{p\big(v_n(t)\mid h_n(t);\sigma_{h_n(t)}\big)}_{v_n(t)\mid h_n(t)\sim\mathcal{CN}(0,\sigma^2_{h_n(t)})}\Big],
\end{eqnarray}
where $p\big(h_n(t);\lambda_{h_n(t)}\big)=\lambda_l$ and $p\big(v_n(t)\mid h_n(t);\sigma_{h_n(t)}\big)$ is the PDF of the $l$th Gaussian component when $h_n(t)=l$.

On the basis of $h_n(t)$, we construct two versions for the complete data of the SAGE algorithm by
\begin{subequations} \label{9}
\begin{eqnarray}
\mathbf{X}_1&=&\big\{\big(\mathbf{x}_n(t),h_n(t)\big)\mid t=1,\dots,T,\nonumber\\
&&n=1,\dots,N\big\},\label{9a}\\
\mathbf{X}_2&=&\big\{\big(y_n(t),h_n(t)\big)\mid t=1,\dots,T,\nonumber\\
&&n=1,\dots,N\big\}, \label{9b}
\end{eqnarray}
\end{subequations}
where $\mathbf{x}_n(t)=[x^1_n(t)~\cdots~x^M_n(t)]^T\in\mathbb{C}^M$ is also missing data introduced and
\begin{equation}  \label{10}  
    y_n(t)=\sum_{m=1}^Mx^m_n(t)
    =\sum_{m=1}^M\Big(\big[\mathbf{a}(\theta_m)s_m(t)\big]_n+v^m_n(t)\Big)
\end{equation}
with $v^1_n(t),\dots,v^M_n(t)$ being \textbf{uncorrelated and identically distributed} and $v_n(t)=\sum_{m=1}^Mv^m_n(t)$. As a consequence, $x^1_n(t),\dots,x^M_n(t)$ are mutually uncorrelated and the complete-data LLFs of both versions are given in \eqref{11} with $\boldsymbol{\Phi}_n(t)=(\boldsymbol{\theta},\mathbf{s}(t),\lambda_{h_n(t)},\sigma_{h_n(t)})$. Obviously, $\mathbf{X}_1$ and $\mathbf{X}_2$ in \eqref{9} let one iteration of the SAGE algorithm consist of multiple EM-pairs.
\begin{figure*}
\begin{subequations}  \label{11}
\begin{eqnarray}  \label{11a}
l(\mathbf{X}_1;\boldsymbol{\Phi})&=&\sum_{t=1}^T\sum_{n=1}^N\ln\Big[ p\big(\mathbf{x}_n(t),h_n(t);\boldsymbol{\Phi}_n(t)\big)\Big]\nonumber\\
&=&\sum_{t=1}^T\sum_{n=1}^N\ln\Big[ p\big(h_n(t);\lambda_{h_n(t)}\big)\prod_{m=1}^M \overbrace{p\big(x^m_n(t)\mid h_n(t);\theta_m,s_m(t),\sigma_{h_n(t)}\big)}^{x^m_n(t)\mid h_n(t)\sim\mathcal{CN}([\mathbf{a}(\theta_m)s_m(t)]_n,\sigma^2_{h_n(t)}/M)}\Big]\nonumber\\
&=&\sum_{t=1}^T\sum_{n=1}^N\Big[\ln\big(\lambda_{h_n(t)}\big)-M\ln\big(\frac{\pi}{M}\big)-M\ln\big(\sigma^2_{h_n(t)}\big)-
\frac{M}{\sigma^2_{h_n(t)}}\sum_{m=1}^M\Big\vert x^m_n(t)-\big[\mathbf{a}(\theta_m)s_m(t)\big]_n\Big\vert^2\Big],\\
l(\mathbf{X}_2;\boldsymbol{\Phi})&=&\sum_{t=1}^T\sum_{n=1}^N\ln\Big[p\big(y_n(t),h_n(t);\boldsymbol{\Phi}_n(t)\big)\Big]=\sum_{t=1}^T\sum_{n=1}^N\ln\Big[p\big(h_n(t);\lambda_{h_n(t)}\big)\times \overbrace{p\big(y_n(t)\mid h_n(t);\boldsymbol{\theta},\mathbf{s}(t),\sigma_{h_n(t)}\big)}^{y_n(t)\mid h_n(t)\sim\mathcal{CN}([\mathbf{A}\mathbf{s}(t)]_n,\sigma^2_{h_n(t)})}\Big]\nonumber\\
&=&\sum_{t=1}^T\sum_{n=1}^N\Big[\ln\big(\lambda_{h_n(t)}\big)-\ln(\pi)-\ln\big(\sigma^2_{h_n(t)}\big)-\frac{1}{\sigma^2_{h_n(t)}}\Big\vert y_n(t)-\big[\mathbf{A}\mathbf{s}(t)\big]_n\Big\vert^2\Big]\label{11b}.
\end{eqnarray}
\end{subequations}
\end{figure*}

\subsection{Procedure}

In this subsection, we elaborate one iteration of the SAGE algorithm, which consists of three EM-pairs. Specifically, 
\begin{itemize}
    \item the first updates $(\hat{\boldsymbol{\theta}},\hat{\mathbf{S}})$ based on $\mathbf{X}_1$,
    \item the second again updates $\hat{\mathbf{S}}$ based on $\mathbf{X}_2$,
    \item the third updates $(\hat{\boldsymbol{\lambda}},\hat{\boldsymbol{\sigma}})$ based on $\mathbf{X}_2$.
\end{itemize}

\subsubsection{E-step of the first EM-pair}

Given $\bar{\boldsymbol{\Phi}}=(\bar{\boldsymbol{\theta}},\bar{\mathbf{S}},\bar{\boldsymbol{\lambda}},\bar{\boldsymbol{\sigma}})$, this step computes the conditional expectation of the complete-data LLF \eqref{11a} by
\begin{eqnarray}  \label{12}
l(\boldsymbol{\Phi};\bar{\boldsymbol{\Phi}})&=&\mathcal{E}\big\{l(\mathbf{X}_1;\boldsymbol{\Phi})\mid\mathbf{Y};\bar{\boldsymbol{\Phi}}\big\}\nonumber\\
&=&C_1-\sum_{t=1}^T\sum_{n=1}^N\sum_{h_n(t)=1}^L\alpha_{h_n(t)}\times\frac{M}{\sigma^2_{h_n(t)}}\times\nonumber\\
&&
\sum_{m=1}^M\Big\vert\mu^m_n(t)-\big[\mathbf{a}(\theta_m)s_m(t)\big]_n\Big\vert^2,
\end{eqnarray}
where $C_1$ denotes an expression unassociated with any parameters in $(\boldsymbol{\theta},\mathbf{S})$, $\alpha_{h_n(t)}$ is the conditional probability of $h_n(t)$ with
\begin{eqnarray}  \label{13}  
\alpha_{h_n(t)}&=&p\big(h_n(t)\mid\mathbf{Y};\bar{\boldsymbol{\Phi}}\big)=p\big(h_n(t)\mid y_n(t);\bar{\boldsymbol{\Phi}}\big)\nonumber\\
&=&\frac{p\big(h_n(t);\bar{\lambda}_{h_n(t)}\big)\times p\big(y_n(t)\mid h_n(t);\bar{\boldsymbol{\Phi}}\big)}{p\big(y_n(t);\bar{\boldsymbol{\Phi}}\big)}\nonumber\\
&=&\frac{\frac{\bar{\lambda}_{h_n(t)}}{\bar{\sigma}^2_{h_n(t)}}\exp\big[-\frac{\vert y_n(t)-[\bar{\mathbf{A}}\bar{\mathbf{s}}(t)]_n\vert^2}{\bar{\sigma}^2_{h_n(t)}}\big]}{\sum_{l=1}^L\frac{\bar{\lambda}_l}{\bar{\sigma}^2_l}\exp\big[-\frac{\vert y_n(t)-[\bar{\mathbf{A}}\bar{\mathbf{s}}(t)]_n\vert^2}{\bar{\sigma}^2_l}\big]}\nonumber\\
    &\in&(0,1),~~~h_n(t)\in\{1,\dots,L\},
\end{eqnarray}
and $\mu^m_n(t)$ is the conditional expectation of $x^m_n(t)$, i.e.,
\begin{eqnarray}  \label{14}  
    &&\mu^m_n(t)=\mathcal{E}\big\{x^m_n(t)\mid\mathbf{Y},h_n(t) ;\bar{\boldsymbol{\Phi}}\big\}\nonumber\\
    &&=\mathcal{E}\big\{x^m_n(t)\mid y_n(t),h_n(t);\bar{\boldsymbol{\Phi}}\big\}\nonumber\\
&&=[\mathbf{a}({\bar{\theta}_m})\bar{s}_m(t)]_n+\big(y_n(t)-[\bar{\mathbf{A}}\bar{\mathbf{s}}(t)]_n\big)/M.
\end{eqnarray}

\subsubsection{M-step of the first EM-pair}

This step updates $(\hat{\boldsymbol{\theta}},\hat{\mathbf{S}})$ by maximizing the expected complete-data LLF \eqref{12} with respect to $(\boldsymbol{\theta},\mathbf{S})$ while holding $(\boldsymbol{\lambda},\boldsymbol{\sigma})=(\bar{\boldsymbol{\lambda}},\bar{\boldsymbol{\sigma}})$ fixed, i.e.,
\begin{eqnarray}  \label{15}  
\min_{\boldsymbol{\theta}\in(0,\pi)^M,\mathbf{S}}\sum_{t=1}^T\sum_{n=1}^N\sum_{l=1}^L\frac{\alpha_{h_n(t)=l}}{\bar{\sigma}^2_{h_n(t)=l}}\times\nonumber\\
\sum_{m=1}^M\Big\vert\mu^m_n(t)-\big[\mathbf{a}(\theta_m)s_m(t)\big]_n\Big\vert^2.
\end{eqnarray}
This problem can be changed to
\begin{eqnarray}  \label{16}  
\min_{\boldsymbol{\theta}\in(0,\pi)^M,\mathbf{S}}\sum_{m=1}^M\sum_{t=1}^T\big[\boldsymbol{\mu}^m(t)-\mathbf{a}(\theta_m)s_m(t)\big]^H\cdot\nonumber\\
\mathbf{H}(t)\big[\boldsymbol{\mu}^m(t)-\mathbf{a}(\theta_m)s_m(t)\big]
\end{eqnarray}
where $\boldsymbol{\mu}^m(t)=[\mu^m_1(t)~\cdots~\mu^m_N(t)]^T\in\mathbb{C}^N$ and
\begin{equation}
    \mathbf{H}(t)=\sum_{l=1}^L\Big(\frac{1}{\bar{\sigma}^2_l}\times\mathrm{diag}\big\{\alpha_{h_1(t)=l},\dots,\alpha_{h_N(t)=l}\big\}\Big)\succ\mathbf{0}_N,\nonumber
\end{equation}
and then can be decomposed into $M$ \textbf{parallel} subproblems based on $m$, i.e.,
\begin{eqnarray}  \label{17}
\min_{\theta_m\in(0,\pi),\mathbf{s}_m}\sum_{t=1}^T\big[\boldsymbol{\mu}^m(t)-\mathbf{a}(\theta_m)s_m(t)\big]^H\cdot\nonumber\\
\mathbf{H}(t)\big[\boldsymbol{\mu}^m(t)-\mathbf{a}(\theta_m)s_m(t)\big], m=1,\dots,M
\end{eqnarray}
where $\mathbf{s}_m=[s_m(1)~\cdots~s_m(T)]$. Utilizing a separable manner, the solutions of these subproblems or the updated $(\hat{\boldsymbol{\theta}},\hat{\mathbf{S}})$ can be obtained by
\begin{subequations}  \label{18}
\begin{eqnarray}     
\breve{\theta}_m&=&\arg\max_{\theta_m\in(0,\pi)}g_m\big(\cos(\theta_m)\big),\label{18a}\nonumber\\
&&m=1,\dots,M,\\
\widetilde{s}_m(t)&=&\mathbf{a}^H(\breve{\theta}_m)\mathbf{H}(t)\boldsymbol{\mu}^m(t)/\chi(t),\nonumber\\
&&m=1,\dots,M,t=1,\dots,T,\label{18b}
\end{eqnarray}
\end{subequations}
where $\chi(t)=\mathbf{a}^H(\theta_m)\mathbf{H}(t)\mathbf{a}(\theta_m)=\mathrm{Tr}\{\mathbf{H}(t)\}>0$ and
\begin{equation}
g_m\big(\cos(\theta_m)\big)=\sum_{t=1}^T\frac{\vert\mathbf{a}^H(\theta_m)\mathbf{H}(t)\boldsymbol{\mu}^m(t)\vert^2}{\chi(t)}.\nonumber
\end{equation}
This M-step and EM-pair end when we obtain the \textbf{partially} updated estimate $\widetilde{\boldsymbol{\Phi}}=(\breve{\boldsymbol{\theta}},\widetilde{\mathbf{S}},\bar{\boldsymbol{\lambda}},\bar{\boldsymbol{\sigma}})$.

\subsubsection{E-step of the second EM-pair}\label{2E}

Given $\widetilde{\boldsymbol{\Phi}}$, this step computes the conditional expectation of the complete-data LLF \eqref{11b} by
\begin{eqnarray}  \label{19}  
&&l(\boldsymbol{\Phi};\widetilde{\boldsymbol{\Phi}})=\mathcal{E}\{l(\mathbf{X}_2;\boldsymbol{\Phi})\mid\mathbf{Y};\widetilde{\boldsymbol{\Phi}}\}\nonumber\\
&=&C_2+\sum_{t=1}^T\sum_{n=1}^N\sum_{h_n(t)=1}^L\beta_{h_n(t)}\Big[\ln\big(\lambda_{h_n(t)}\big)-\nonumber\\
&&\ln\big(\sigma^2_{h(t)}\big)-\frac{1}{\sigma^2_{h_n(t)}}\Big\vert y_n(t)-\big[\mathbf{A}\mathbf{s}(t)\big]_n\Big\vert^2\Big],
\end{eqnarray}
where $C_2$ is a constant and $\beta_{h_n(t)}$ is the conditional probability of $h_n(t)$ with
\begin{eqnarray}   \label{20}  
\beta_{h_n(t)}&=&p\big(h_n(t)\mid\mathbf{Y};\widetilde{\boldsymbol{\Phi}}\big)=p\big(h_n(t)\mid y_n(t);\widetilde{\boldsymbol{\Phi}}\big)\nonumber\\
&=&\frac{p\big(h_n(t);\bar{\lambda}_{h_n(t)}\big)\times p\big(y_n(t)\mid h_n(t);\widetilde{\boldsymbol{\Phi}}\big)}{p\big(y_n(t);\widetilde{\boldsymbol{\Phi}}\big)}\nonumber\\
&=&\frac{\frac{\bar{\lambda}_{h_n(t)}}{\bar{\sigma}^2_{h_n(t)}}\exp\big[-\frac{\vert y_n(t)-[\breve{\mathbf{A}}\widetilde{\mathbf{s}}(t)]_n\vert^2}{\bar{\sigma}^2_{h_n(t)}}\big]}{\sum_{l=1}^L\frac{\bar{\lambda}_l}{\bar{\sigma}^2_l}\exp\big[-\frac{\vert y_n(t)-[\breve{\mathbf{A}}\widetilde{\mathbf{s}}(t)]_n\vert^2}{\bar{\sigma}^2_l}\big]}\nonumber\\
    &\in&(0,1),~~~h_n(t)\in\{1,\dots,L\}.
\end{eqnarray}

\subsubsection{M-step of the second EM-pair} \label{2M}

This step again updates $\hat{\mathbf{S}}$ by maximizing the expected complete-data LLF \eqref{19} with respect to $\mathbf{S}$ while holding $(\boldsymbol{\theta}, \boldsymbol{\lambda},\boldsymbol{\sigma})=(\breve{\boldsymbol{\theta}},\bar{\boldsymbol{\lambda}},\bar{\boldsymbol{\sigma}})$ fixed, i.e.,
\begin{equation}  \label{21} 
\min_{\mathbf{S}}\sum_{t=1}^T\sum_{n=1}^N\sum_{l=1}^L\frac{\beta_{h_n(t)=l}}{\bar{\sigma}^2_{h_n(t)=l}}\Big\vert y_n(t)-\big[\breve{\mathbf{A}}\mathbf{s}(t)\big]_n\Big\vert^2.
\end{equation}
This problem can be changed to
\begin{equation}  \label{22}  
\min_{\mathbf{S}}\sum_{t=1}^T\big[\mathbf{y}(t)-\breve{\mathbf{A}}\mathbf{s}(t)\big]^H
\mathbf{D}(t)\big[\mathbf{y}(t)-\breve{\mathbf{A}}\mathbf{s}(t)\big]
\end{equation}
where
\begin{equation}
    \mathbf{D}(t)=\sum_{l=1}^L\Big(\frac{1}{\bar{\sigma}^2_l}\times\mathrm{diag}\big\{\beta_{h_1(t)=l},\dots,\beta_{h_N(t)=l}\big\}\Big)\succ\mathbf{0}_N.\nonumber
\end{equation}
The solution or the updated $\hat{\mathbf{S}}$ can be obtained by
\begin{equation}  \label{23}  
\breve{\mathbf{s}}(t)=\big(\breve{\mathbf{A}}^H\mathbf{D}(t)\breve{\mathbf{A}}\big)^{-1}\breve{\mathbf{A}}^H\mathbf{D}(t)\mathbf{y}(t),t=1,\dots,T.
\end{equation}
This M-step and EM-pair end when we obtain the \textbf{partially} updated estimate $\ddot{\boldsymbol{\Phi}}=(\breve{\boldsymbol{\theta}},\breve{\mathbf{S}},\bar{\boldsymbol{\lambda}},\bar{\boldsymbol{\sigma}})$.

\subsubsection{E-step of the third EM-pair}\label{3E}

Given $\ddot{\boldsymbol{\Phi}}$, this step again computes the conditional expectation of the complete-data LLF \eqref{11b} by
\begin{eqnarray}  \label{24}
&&l(\boldsymbol{\Phi};\ddot{\boldsymbol{\Phi}})=\mathcal{E}\{l(\mathbf{X}_2;\boldsymbol{\Phi})\mid\mathbf{Y};\ddot{\boldsymbol{\Phi}}\}\nonumber\\
&=&C_2+\sum_{t=1}^T\sum_{n=1}^N\sum_{h_n(t)=1}^L\xi_{h_n(t)}\Big[\ln\big(\lambda_{h_n(t)}\big)-\nonumber\\
&&\ln\big(\sigma^2_{h(t)}\big)-\frac{1}{\sigma^2_{h_n(t)}}\Big\vert y_n(t)-\big[\mathbf{A}\mathbf{s}(t)\big]_n\Big\vert^2\Big],
\end{eqnarray}
where 
$\xi_{h_n(t)}$ is the conditional probability of $h_n(t)$ with
\begin{eqnarray}  \label{25}  
\xi_{h_n(t)}&=&p\big(h_n(t)\mid\mathbf{Y};\ddot{\boldsymbol{\Phi}}\big)=p\big(h_n(t)\mid y_n(t);\ddot{\boldsymbol{\Phi}}\big)\nonumber\\
&=&\frac{p\big(h_n(t);\bar{\lambda}_{h_n(t)}\big)\times p\big(y_n(t)\mid h_n(t);\ddot{\boldsymbol{\Phi}}\big)}{p\big(y_n(t);\ddot{\boldsymbol{\Phi}}\big)}\nonumber\\
&=&\frac{\frac{\bar{\lambda}_{h_n(t)}}{\bar{\sigma}^2_{h_n(t)}}\exp\big[-\frac{\vert y_n(t)-[\breve{\mathbf{A}}\breve{\mathbf{s}}(t)]_n\vert^2}{\bar{\sigma}^2_{h_n(t)}}\big]}{\sum_{l=1}^L\frac{\bar{\lambda}_l}{\bar{\sigma}^2_l}\exp\big[-\frac{\vert y_n(t)-[\breve{\mathbf{A}}\breve{\mathbf{s}}(t)]_n\vert^2}{\bar{\sigma}^2_l}\big]}\nonumber\\
    &\in&(0,1),~~~h_n(t)\in\{1,\dots,L\}.
\end{eqnarray}

\subsubsection{M-step of the third EM-pair}\label{3M}

This step updates $(\hat{\boldsymbol{\lambda}},\hat{\boldsymbol{\sigma}})$ by maximizing the expected complete-data LLF \eqref{24} with respect to $(\boldsymbol{\lambda},\boldsymbol{\sigma})$ while holding $(\boldsymbol{\theta},\mathbf{S})=(\breve{\boldsymbol{\theta}},\breve{\mathbf{S}})$ fixed, i.e.,
\begin{eqnarray}  \label{26}  
\max_{\boldsymbol{\lambda}>\mathbf{0},\boldsymbol{\lambda}^T\mathbf{1}=1,\boldsymbol{\sigma}>\mathbf{0}}\sum_{t=1}^T\sum_{n=1}^N\sum_{l=1}^L\xi_{h_n(t)=l}\Big[\ln\big(\lambda_{h_n(t)=l}\big)\nonumber\\
-\ln\big(\sigma^2_{h_n(t)=l}\big)-\frac{c_n(t)}{\sigma^2_{h_n(t)=l}}\Big],
\end{eqnarray}
where $c_n(t)=\big\vert y_n(t)-[\breve{\mathbf{A}}\breve{\mathbf{s}}(t)]_n\big\vert^2\ge0$.
This problem can be decomposed into the two \textbf{parallel} subproblems:
\begin{subequations} \label{27}
\begin{eqnarray}  \label{27a}  
&&\max_{\boldsymbol{\lambda}>\mathbf{0},\boldsymbol{\lambda}^T\mathbf{1}=1}\sum_{l=1}^L\ln(\lambda_l)\kappa_l,\\
&&\min_{\boldsymbol{\sigma}>\mathbf{0}}\sum_{l=1}^L\big[
\kappa_l\ln(\sigma^2_l)+\varrho_l/\sigma^2_l\big],\label{27b}
\end{eqnarray}
\end{subequations}
where $\varrho_l=\sum_{t=1}^T\sum_{n=1}^N\big[\xi_{h_n(t)=l}\times c_n(t)\big]>0$ and $\kappa_l=\sum_{t=1}^T\sum_{n=1}^N\xi_{h_n(t)=l}>0$ with $\sum_{l=1}^L\kappa_l=TN$. Using the Lagrange multiplier method \cite{Boyd}, the solution of problem \eqref{27a} or the updated $\hat{\boldsymbol{\lambda}}$ can be obtained by
\begin{eqnarray} \label{28}
\breve{\lambda}_l=\kappa_l/(TN)>0,l=1,\dots,L.
\end{eqnarray}
Moreover, problem \eqref{27b} can be decomposed into the $L$ \textbf{parallel} subproblems:
\begin{eqnarray}  \label{29}  
\min_{\sigma_l>0}
\kappa_l\ln(\sigma^2_l)+\varrho_l/\sigma^2_l,l=1,\dots,L.
\end{eqnarray}
The solutions of these subproblems or the updated $\hat{\boldsymbol{\sigma}}$ can be obtained by
\begin{eqnarray} \label{30}
\breve{\sigma}_l=\sqrt{\varrho_l/\kappa_l}>0,l=1,\dots,L.
\end{eqnarray}
This M-step, EM-pair, and iteration end when we obtain the \textbf{fully} updated estimate $\breve{\boldsymbol{\Phi}}=(\breve{\boldsymbol{\theta}},\breve{\mathbf{S}},\breve{\boldsymbol{\lambda}},\breve{\boldsymbol{\sigma}})$. The straightforward procedure of this SAGE algorithm is provided in \textbf{Algorithm \ref{alg:A}}.

\begin{algorithm}
\caption{SAGE Algorithm} \label{alg:A}
\begin{algorithmic}[1]
\STATE {Initialize $\hat{\boldsymbol{\Phi}}=(\hat{\boldsymbol{\theta}},\hat{\mathbf{S}},\hat{\boldsymbol{\lambda}},\hat{\boldsymbol{\sigma}})$, $k=1$, and $K$.}
        \WHILE{$k\le K$}
           \STATE {Update $\hat{\boldsymbol{\theta}}$ and $\hat{\mathbf{S}}$ using \eqref{18a} (or \textbf{Algorithm \ref{alg:D}}) and \eqref{18b}, respectively.}
           \STATE {Again update $\hat{\textbf{S}}$ using \eqref{23}.}
           \STATE {Update $\hat{\boldsymbol{\lambda}}$ and $\hat{\boldsymbol{\sigma}}$ using \eqref{28} and \eqref{30}, respectively.}
           \STATE{$k=k+1$.}
       \ENDWHILE
\STATE {Output $\hat{\boldsymbol{\theta}}$.}
\end{algorithmic}
\end{algorithm}

\section{AECM Algorithm}

The SAGE algorithm designed by Kozick and Sadler only utilizes the more informative complete data $\mathbf{X}_1$ in \eqref{9a} to simultaneously update all of the $\hat{\theta}_m$'s by \eqref{18a} at each iteration, which cannot yield the fastest convergence \cite{Fessler,Gong}. More importantly, the SAGE algorithm cannot properly converge under unequal signal powers when using Eq. \eqref{18a} to update the $\hat{\theta}$'s at each iteration. These motivate the AECM algorithm designed in this section.

\subsection{Complete Data}

To design the AECM algorithm with faster stable convergence, we construct $M+1$ less informative versions for the complete data of the AECM algorithm by 
\begin{subequations}  \label{31}  
\begin{eqnarray}
\mathbf{Z}_1&=&\big\{\big(z^1_n(t),h_n(t)\big)\mid t=1,\dots,T,\nonumber\\
&&n=1,\dots,N\big\},\\
&&~~~\vdots~~~~~~~~~~~\vdots~~~~~~~~~~~\vdots\nonumber\\
\mathbf{Z}_M&=&\big\{\big(z^M_n(t),h_n(t)\big)\mid t=1,\dots,T,\nonumber\\
&&n=1,\dots,N\big\},\\
\mathbf{X}_2&=&\big\{\big(y_n(t),h_n(t)\big)\mid t=1,\dots,T,\nonumber\\
&&n=1,\dots,N\big\},
\end{eqnarray}
\end{subequations}
where $z^m_n(t)=[\mathbf{a}(\theta_m)s_m(t)]_n+v_n(t)$ is new missing data introduced and different from $x^m_n(t)$ in \eqref{10}, i.e.,
\begin{equation}  \label{32}
y_n(t)=\underbrace{z^m_n(t)}_{\mathrm{random}}+\sum_{d\ne m}\underbrace{\big[\mathbf{a}(\theta_d)s_d(t)\big]_n}_{\mathrm{deterministic}}.
\end{equation}
These $M+1$ versions in \eqref{31} give rise to one iteration consisting of $M+2$ EM-cycles.

\subsection{Procedure}

In this subsection, we describe one iteration of the AECM algorithm, which consists of $M+2$ EM-cycles and updates the $\hat{\theta}_m$'s sequentially (one by one). Specifically,
\begin{itemize}
    \item the $m$th $(m=1,\dots,M)$ updates $(\hat{\theta}_m,\hat{\mathbf{s}}_m)$ based on $\mathbf{Z}_m$,
    \item the $(M+1)$th again updates $\hat{\mathbf{S}}$ based on $\mathbf{X}_2$,
    \item the $(M+2)$th updates $(\hat{\boldsymbol{\lambda}},\hat{\boldsymbol{\sigma}})$ based on $\mathbf{X}_2$.
\end{itemize}

Since the $(M+1)$th and $(M+2)$th EM-cycles are the same as the second and third EM-pairs in the SAGE algorithm, we only elaborate the $m$th $(m=1,\dots,M)$ EM-cycle composed of one E-step and one CM-step below. For notational convenience, let $\theta^{(m)}$ be the updated $\hat{\theta}$ obtained at the $m$th EM-cycle, $\hat{\boldsymbol{\phi}}_m=(\hat{\theta}_m,\hat{\mathbf{s}}_m)$, and 
\begin{eqnarray}
    \mathbf{A}^{(m-1)}&=&\big[\mathbf{a}(\breve{\theta}_1)~\cdots~\mathbf{a}(\breve{\theta}_{m-1})~\mathbf{a}(\bar{\theta}_m)~\cdots~\mathbf{a}(\bar{\theta}_M)\big],\nonumber\\
\mathbf{s}^{(m-1)}(t)&=&\big[\widetilde{s}_1(t)~\cdots~\widetilde{s}_{m-1}(t)~\bar{s}_m(t)~\cdots~\bar{s}_M(t)\big]^T,\nonumber\\
    \boldsymbol{\Phi}^{(m-1)}&=&(\widetilde{\boldsymbol{\phi}}_1\dots,\widetilde{\boldsymbol{\phi}}_{m-1},\bar{\boldsymbol{\phi}}_m,\dots,\bar{\boldsymbol{\phi}}_M,\bar{\boldsymbol{\lambda}},\bar{\boldsymbol{\sigma}}),\nonumber\\
    \widetilde{\boldsymbol{\phi}}_{m-1}&=&(\breve{\theta}_{m-1},\widetilde{\mathbf{s}}_{m-1}),\bar{\boldsymbol{\phi}}_m=(\bar{\theta}_m,\bar{\mathbf{s}}_m).\nonumber
\end{eqnarray}

\subsubsection{E-step of the $m$th EM-cycle}

Following \eqref{11}, we first write the complete-data LLF of $\mathbf{Z}_m$ in \eqref{31} as
\begin{eqnarray}  \label{33}
l(\mathbf{Z}_m;\boldsymbol{\Phi})=\sum_{t=1}^T\sum_{n=1}^N\ln\Big[p\big(h_n(t);\lambda_{h_n(t)}\big)\times\nonumber\\
p\big(z^m_n(t)\mid h_n(t);\theta_m,s_m(t),\sigma_{h_n(t)}\big)\Big]\nonumber\\
=\sum_{t=1}^T\sum_{n=1}^N\big[\ln\big(\lambda_{h_n(t)}\big)-\ln(\pi)-\ln\big(\sigma^2_{h_n(t)}\big)-\nonumber\\
\frac{1}{\sigma^2_{h_n(t)}}\Big\vert z^m_n(t)-\big[\mathbf{a}(\theta_m)s_m(t)\big]_n\Big\vert^2\Big].
\end{eqnarray}
Then, given $\boldsymbol{\Phi}^{(m-1)}$, this step computes its conditional expectation by
\begin{eqnarray}  \label{34}
l(\boldsymbol{\Phi};\boldsymbol{\Phi}^{(m-1)})=\mathcal{E}\big\{l(\mathbf{Z}_m;\boldsymbol{\Phi})\mid\mathbf{Y};\boldsymbol{\Phi}^{(m-1)}\big\}=C_2\nonumber\\
+\sum_{t=1}^T\sum_{n=1}^N\sum_{h_n(t)=1}^L\delta^{(m)}_{h_n(t)}\Big[\ln\big(\lambda_{h_n(t)}\big)-\ln\big(\sigma^2_{h_n(t)}\big)\nonumber\\
-\frac{1}{\sigma^2_{h_n(t)}}\Big(\rho^m_n(t)+\Big\vert \eta^m_n(t)-\big[\mathbf{a}(\theta_m)s_m(t)\big]_n\Big\vert^2\Big)\Big],
\end{eqnarray}
where $\delta^{(m)}_{h_n(t)}$ is the conditional probability of $h_n(t)$ with
\begin{eqnarray}  \label{35}  
&&\delta^{(m)}_{h_n(t)}=p\big(h_n(t)\mid y_n(t);\boldsymbol{\Phi}^{(m-1)}\big)\nonumber\\
&&=\frac{p\big(h_n(t);\bar{\lambda}_{h_n(t)}\big)\times p\big(y_n(t)\mid h_n(t);\boldsymbol{\Phi}^{(m-1)}\big)}{p\big(y_n(t);\boldsymbol{\Phi}^{(m-1)}\big)}\nonumber\\
&&=\frac{\frac{\bar{\lambda}_{h_n(t)}}{\bar{\sigma}_{h_n(t)}^2}\exp\big[-\frac{\vert y_n(t)-[\mathbf{A}^{(m-1)}\mathbf{s}^{(m-1)}(t)]_n\vert^2}{\bar{\sigma}_{h_n(t)}^2}\big]}{\sum_{l=1}^L\frac{\bar{\lambda}_l}{\bar{\sigma}_l^2}\exp\big[\frac{\vert y_n(t)-[\mathbf{A}^{(m-1)}\mathbf{s}^{(m-1)}(t)]_n\vert^2}{-\bar{\sigma}_l^2}\big]}\nonumber\\
    &&\in(0,1),~~~h_n(t)\in\{1,\dots,L\},
\end{eqnarray}
$\eta^m_n(t)$ and $\rho^m_n(t)$ are, respectively, the conditional expectation and variance of $z^m_n(t)$, i.e.,
\begin{subequations}
\begin{eqnarray}   \label{36}
    \eta^m_n(t)=\mathcal{E}\big\{z^m_n(t)\mid\mathbf{Y},h_n(t);\boldsymbol{\Phi}^{(m-1)}\big\}
    \nonumber\\
=y_n(t)-\sum_{d<m}\big[\mathbf{a}(\breve{\theta}_d)\widetilde{s}_d(t)\big]_n-\sum_{d>m}\big[\mathbf{a}(\bar{\theta}_d)\bar{s}_d(t)\big]_n,\\
\rho^m_n(t)=\mathcal{D}\big\{z^m_n(t)\mid\mathbf{Y},h_n(t);\boldsymbol{\Phi}^{(m-1)}\big\}=0.
\end{eqnarray}
\end{subequations}

\subsubsection{CM-step of the $m$th EM-cycle}

This step tries updating $(\hat{\theta}_m,\hat{\mathbf{s}}_m)$ by maximizing the expected complete-data LLF \eqref{34} with respect to $(\theta_m,\mathbf{s}_m)$ while holding $(\boldsymbol{\lambda},\boldsymbol{\sigma})=(\bar{\boldsymbol{\lambda}},\bar{\boldsymbol{\sigma}})$ fixed, i.e.,
\begin{eqnarray}  \label{37} 
\min_{\theta_m\in(0,\pi),\mathbf{s}_m}\sum_{t=1}^T\sum_{n=1}^N\sum_{l=1}^L\frac{\delta^{(m)}_{h_n(t)=l}}{\bar{\sigma}_{h_n(t)=l}^2}\times\nonumber\\
\Big\vert \eta_n^m(t)-\big[\mathbf{a}(\theta_m)s_m(t)\big]_n\Big\vert^2.
\end{eqnarray} 
This problem can be changed to
\begin{eqnarray}  \label{38}  
\min_{\theta_m\in(0,\pi),\mathbf{s}_m}\sum_{t=1}^T\big[\boldsymbol{\eta}^m(t)-\mathbf{a}(\theta_m)s_m(t)\big]^H
\mathbf{G}^{(m)}(t)\cdot\nonumber\\
\big[\boldsymbol{\eta}^m(t)-\mathbf{a}(\theta_m)s_m(t)\big]
\end{eqnarray}
where $\boldsymbol{\eta}^m(t)=[\eta^m_1(t)~\cdots~\eta^m_N(t)]^T$ and 
\begin{eqnarray}
    \mathbf{G}^{(m)}(t)&=&\sum_{l=1}^L\Big(\frac{1}{\bar{\sigma}_l^2}\times\nonumber\\
    &&\mathrm{diag}\big\{\delta^{(m)}_{h_1(t)=l},\dots,\delta^{(m)}_{h_N(t)=l}\big\}\Big)\succ\mathbf{0}_N.\nonumber
\end{eqnarray}
Following subproblems \eqref{17}, the solution or the updated $(\hat{\theta}_m,\hat{\mathbf{s}}_m)$ can be obtained by
\begin{subequations} \label{39}
\begin{eqnarray}     %
\breve{\theta}_m=\arg\max_{\theta_m\in(0,\pi)}d_m\big(\cos(\theta_m)\big),\label{39a}\\
\widetilde{s}_m(t)=\frac{\mathbf{a}^H(\breve{\theta}_m)\mathbf{G}^{(m)}(t)\boldsymbol{\eta}^m(t)}{\chi^{(m)}(t)},t=1,\dots,T,\label{39b}
\end{eqnarray}
\end{subequations}
where $\chi^{(m)}(t)=\mathrm{Tr}\{\mathbf{G}^{(m)}(t)\}>0$ and
\begin{equation}
d_m\big(\cos(\theta_m)\big)=\sum_{t=1}^T\frac{\vert\mathbf{a}^H(\theta_m)\mathbf{G}^{(m)}(t)\boldsymbol{\eta}^m(t)\vert^2}{\chi^{(m)}(t)}.\nonumber
\end{equation}

Unfortunately, Fig. \ref{f0} illustrates that under unequal signal powers, i.e., $P_1=\cdots=P_M$ is not satisfied, the AECM algorithm cannot also properly converge when using \eqref{39a} to update the $\hat{\theta}_m$'s at each iteration and all of the $\hat{\theta}_m$'s obtained get close to the true DOA of the source with the largest/larger signal power. In order for addressing this issue, we resort to the golden section search method\footnote{We do not adopt the gradient ascent method since simulation results show that this method easily gets stuck in an endless loop due to finite precision when searching the local maximum point of $d_m\big(\cos(\theta_m)\big)$ or $g_m\big(\cos(\theta_m)\big)$ closest to $\bar{\theta}_m$.} in \textbf{Algorithm \ref{alg:D}} to find the local (not global) maximum point of $d_m\big(\cos(\theta_m)\big)$ in \eqref{39a} or $g_m\big(\cos(\theta_m)\big)$ in \eqref{18a} closest to $\bar{\theta}_m$ as $\breve{\theta}_m$, which can reduce the difference between $\bar{\theta}_m$ and $\breve{\theta}_m$ at early iterations and guarantee the monotonicity. This CM-step and EM-cycle end when we obtain the \textbf{partially} updated estimate $\boldsymbol{\Phi}^{(m)}$.

\begin{remark}
    \textbf{Algorithm \ref{alg:D}} indicates that the AECM algorithm does not require maximizing $d_m\big(\cos(\theta_m)\big)$ and the expected complete-data LLF \eqref{34} with respect to $(\theta_m,\mathbf{s}_m)$, which lets the algorithm be the AECM algorithm \cite{Meng}, instead of the SAGE algorithm \cite{Fessler}. Accordingly, the SAGE algorithm designed by Kozick and Sadler in \cite{Kozick} also becomes the AECM algorithm when using \textbf{Algorithm \ref{alg:D}} to update the $\hat{\theta}_m$'s at each iteration, but we still call it ``the SAGE algorithm'' for convenience.
\end{remark}

\begin{algorithm}
\caption{Golden Section Based DOA Search} \label{alg:D}
\begin{algorithmic}[1]
\STATE {Let $f_m(u)=g_m(u)$ in \eqref{18a}\\ or ~~$f_m(u)=d_m(u)$ in \eqref{39a}.}
\STATE {Initialize $\bar{u}=\cos(\bar{\theta}_m)$, $\Delta u>0$, $\epsilon\in(0,2\Delta u)$, and $s=1$.}
\IF{$f'_m(\bar{u})<0$}
\STATE{$s=-1$.}
\ENDIF
\STATE{$\breve{u}=\bar{u}+s\times\Delta u$.}
\WHILE{$f_m(\bar{u})<f_m(\breve{u})$}
    \STATE {$\bar{u}=\breve{u}$ and $\breve{u}=\breve{u}+s\times\Delta u$.}
\ENDWHILE
\IF{$s=1$}
    \STATE{$start=\bar{u}-\Delta u$ and $end=\breve{u}$.}
\ELSE
    \STATE{$start=\breve{u}$ and $end=\bar{u}+\Delta u$.}
\ENDIF
\WHILE{$end-start>\epsilon$}
\STATE {$mid_1=start+0.382(end-start)$ and $mid_2=start+0.618(end-start)$.}
\IF{$f_m(mid_1)<f_m(mid_2)$}
   \STATE{$start=mid_1$.}
\ELSE
   \STATE{$end=mid_2$.}
\ENDIF
\ENDWHILE
\STATE{$\breve{u}=(start+end)/2$ and $\breve{\theta}_m=\arccos(\breve{u})$.}
\end{algorithmic}
\end{algorithm}

\begin{figure} 
\centerline{\includegraphics[width=21pc]{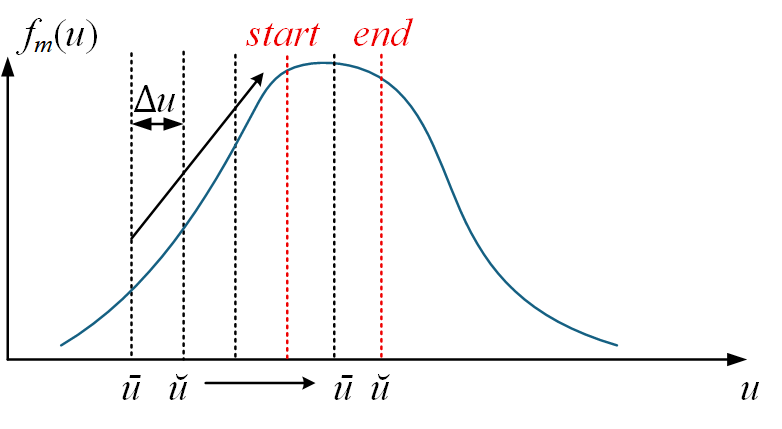}}
\caption{Diagram of searching the initial unimodal interval $[start,end]$.}\label{gss}
\end{figure}

\begin{remark}
In \textbf{Algorithm \ref{alg:D}}, lines 3-14 are designed to search the initial unimodal interval $[start,end]$, as shown in Fig. \ref{gss}. Obviously, the search ``resolution'' $\Delta u$ should be as small as possible, which, however, makes the algorithm computationally expensive. Moreover, due to the length of the initial unimodal interval $end-start=2\Delta u$, the tolerance parameter $\epsilon$ must be less than $2\Delta u$, i.e., $\epsilon<2\Delta u$.
\end{remark}

After the $M+2$ EM-cycle, we obtain the \textbf{fully} updated estimate $\breve{\boldsymbol{\Phi}}=(\breve{\boldsymbol{\theta}},\breve{\mathbf{S}},\breve{\boldsymbol{\lambda}},\breve{\boldsymbol{\sigma}})$. The straightforward procedure of this AECM algorithm is provided in \textbf{Algorithm \ref{alg:B}}.

\begin{algorithm}
\caption{AECM Algorithm} \label{alg:B}
\begin{algorithmic}[1]
\STATE {Initialize $\hat{\boldsymbol{\Phi}}=(\hat{\boldsymbol{\phi}}_1,\dots,\hat{\boldsymbol{\phi}}_M,\hat{\boldsymbol{\lambda}},\hat{\boldsymbol{\sigma}})$, $k=1$, and $K$.}
\WHILE{$k\le K$}
    \FOR{$m=1,\dots,M$}
    \STATE {Update $(\hat{\theta}_m,\hat{\mathbf{s}}_m)$ using \textbf{Algorithm \ref{alg:D}} (or \eqref{39a}) and \eqref{39b}.} 
    \ENDFOR
    \STATE {Again update $\hat{\textbf{S}}$ using \eqref{23}.}
           \STATE {Update $\hat{\boldsymbol{\lambda}}$ and $\hat{\boldsymbol{\sigma}}$ using \eqref{28} and \eqref{30}, respectively.}
    \STATE {$k=k+1$.}
\ENDWHILE
\STATE {Output $\hat{\boldsymbol{\theta}}$.}
\end{algorithmic}
\end{algorithm}

\subsection{Convergence and Complexity}

According to \cite{GMcLachlan,Meng,Wu}, it can be verified that the AECM algorithm meets certain ``regularity'' conditions and its sequence of parameter estimates always converges to a specific point of $f(\boldsymbol{\Phi})$. That is to say, the AECM algorithm converges to the global optimum of problem \eqref{7} when the initialization or initial point is good. Moreover, the AECM algorithm updates the $\hat{\theta}_m$'s sequentially (one by one) at each iteration and thus yields faster convergence than the SAGE algorithm, which will be shown in Fig. \ref{f1}.

From TABLE \ref{t1} and TABLE \ref{t2}, we can notice that the computational complexities of each iteration in both algorithms are dominated by updating the $\hat{\theta}_m$'s when $T>N>M$ and $L$ is small, so the AECM algorithm possesses nearly the same computational complexity of each iteration as the SAGE algorithm when both algorithms use \textbf{Algorithm \ref{alg:D}} to update the $\hat{\theta}_m$'s at each iteration. Accordingly, the AECM algorithm is more attractive in computation due to the faster convergence.

\begin{table*}
    \caption{Calculation burden of one iteration in the SAGE algorithm}
    \centering
    \begin{tabular}{c | c | c | c | c | c}
    \hline 
    \textbf{Update} & \makecell{$\alpha_{h_n(t)}$'s, $\beta_{h_n(t)}$'s,\\and $\xi_{h_n(t)}$'s} & $\hat{\theta}_m$'s & $\hat{\mathbf{S}}$ & $\kappa_l$'s & $\varrho_l$'s \\
    \hline
    \textbf{Method} & \makecell{explicit formulas \eqref{13}, \eqref{20},\\and \eqref{25}} & \makecell{$M$ one-dimensional\\searches} & \makecell{explicit formulas \eqref{18b}\\and \eqref{23}}  & explicit formula \eqref{27} & explicit formula \eqref{27} \\
    \hline 
    \textbf{Complexity} & $\mathit{\mathcal{O}}(TNL)$ &  $\mathit{\mathcal{O}}\big(\frac{\ln(\epsilon/(2\Delta u)}{\ln(0.618)}MTN^2\big)$ &  $\mathit{\mathcal{O}}(MTN^2)$ &  $\mathit{\mathcal{O}}(TNL)$ & $\mathit{\mathcal{O}}(TNL)$ \\
    \hline 
    \end{tabular}
    \label{t1}
\vspace{0.1cm}
    \caption{Calculation burden of one iteration in the AECM algorithm}
    \centering
    \begin{tabular}{c | c | c | c | c | c}
    \hline 
    \textbf{Update} & $\delta^{(m)}_{h_n(t)}$'s & $\hat{\theta}_m$'s & $\hat{\mathbf{S}}$ & $\kappa_l$'s & $\varrho_l$'s \\
    \hline
    \textbf{Method} & explicit formula \eqref{35} & \makecell{$M$ one-dimensional searches} & \makecell{explicit formulas \eqref{39b}\\and \eqref{23}}  & explicit formula \eqref{27} & explicit formula \eqref{27} \\
    \hline 
    \textbf{Complexity} & $\mathit{\mathcal{O}}(MTNL)$ &  $\mathit{\mathcal{O}}\big(\frac{\ln(\epsilon/(2\Delta u)}{\ln(0.618)}MTN^2\big)$ &  $\mathit{\mathcal{O}}(MTN^2)$ &  $\mathit{\mathcal{O}}(TNL)$ & $\mathit{\mathcal{O}}(TNL)$ \\
    \hline 
    \end{tabular}
    \label{t2}
\end{table*}

\section{Numerical Results}

In this section, simulation results are used to compare and prove the SAGE algorithm and the AECM algorithm. Unless otherwise specified in the following figures, let $N=6$, $M=2$, $\theta_1=60^{\circ}$, $\theta_2=100^{\circ}$, $T=200$, $s_1(t)=1$, $s_2(t)=\sqrt{10}$. The impulsive noise parameters are $L=2$, $\lambda_1=0.95$, $\lambda_2=0.05$, $\sigma_1=1$, $\sigma_2=\sqrt{20}$ \cite{Kozick}. Moreover, initialize $\hat{\theta}_1=55^{\circ}$, $\hat{\theta}_2=105^{\circ}$, $\hat{s}_1(t)=\hat{s}_2(t)=1$, $\hat{\lambda}_1=0.9$, $\hat{\lambda}_2=0.1$, $\hat{\sigma}_1=1$, $\hat{\sigma}_2=\sqrt{10}$, $K=50$, $\Delta u=10^{-3}$, and $\epsilon=10^{-4}$.
\begin{remark}
    $s_1(t)=1$ and $s_2(t)=\sqrt{10}$ indicate $P_1=1$, $P_2=10$, and $P_1<P_2$. More importantly, $s_2(t)=\sqrt{10}\times s_1(t)$, i.e., both signals are coherent\cite{Krim}.
\end{remark}

\subsection{Convergence Comparison}

\begin{figure}
\centerline{\includegraphics[width=22pc]{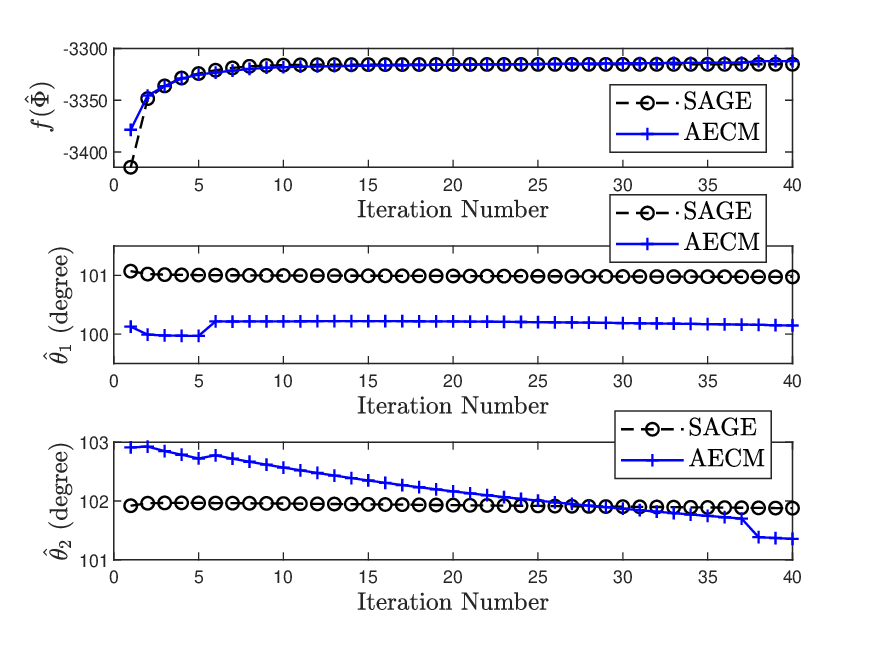}}
\caption{Convergence comparison of both algorithms not using \textbf{Algorithm 2}.}\label{f0}
\end{figure}

\begin{figure} 
\centerline{\includegraphics[width=22pc]{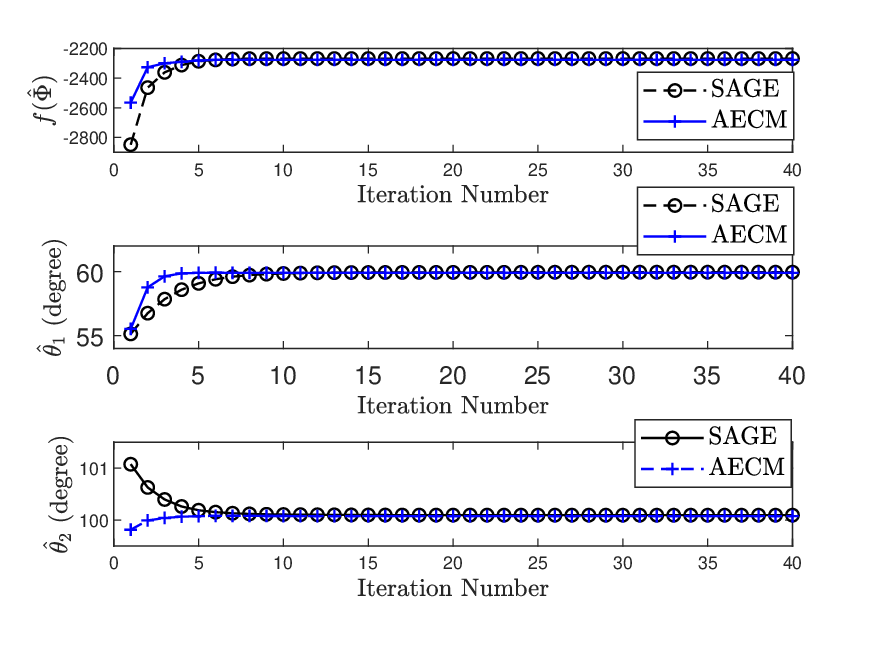}}
\caption{Convergence comparison of both algorithms using \textbf{Algorithm 2}.}\label{f1}
\end{figure}

Fig. \ref{f0} compares both algorithms when both algorithms, respectively, employ \eqref{18a} and \eqref{39a} to update the $\hat{\theta}_m$'s at each iteration given the same $\mathbf{Y}$\footnote{This can be achieved by first finding a grid point closest to the global maximum point of $f_m(u)$ as $\bar{u}$ for initialization in \textbf{Algorithm \ref{alg:D}}.}. We can observe that under $P_1<P_2$, both algorithms improperly converge although the initialization is good, namely, the sequences of $\hat{\theta}_1$ and $\hat{\theta}_2$ from both algorithms get close to the true DOA $\theta_2=100^{\circ}$ with the higher signal power. On the contrary, Fig. \ref{f1} compares both algorithms when both algorithms employ \textbf{Algorithm \ref{alg:D}} to update the $\hat{\theta}_m$'s at each iteration given the same $\mathbf{Y}$. We can observe that both algorithms properly converge when the initialization is good, and the AECM algorithm exhibits faster stable converge than the SAGE algorithm. Because the AECM algorithm possesses nearly the same computational complexity of each iteration as the SAGE algorithm, the AECM algorithm is more attractive in computation.

\begin{figure} 
\centerline{\includegraphics[width=20pc]{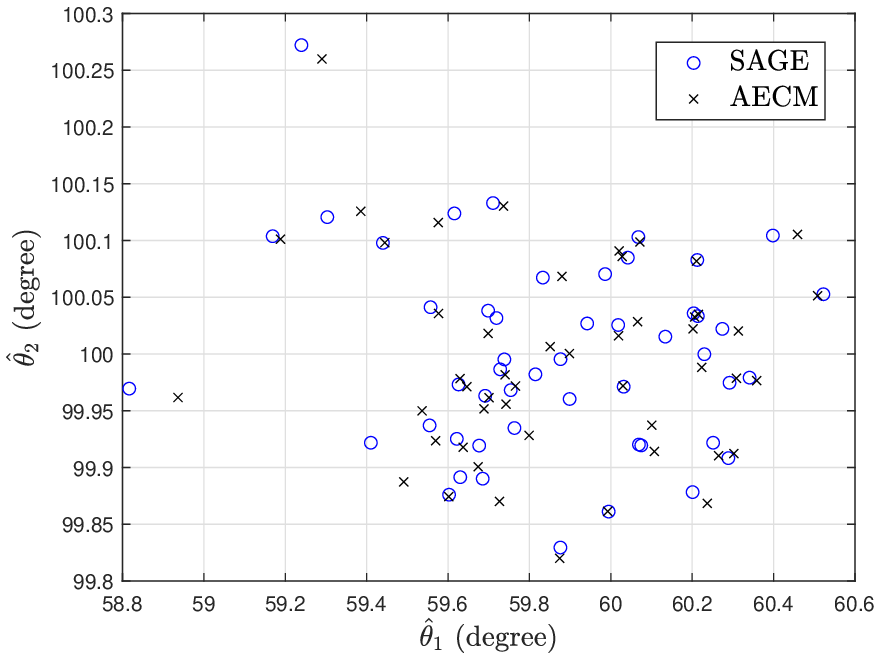}}
\caption{Scatter plot of $(\hat{\theta}_1,\hat{\theta}_2)$’s from both algorithms using \textbf{Algorithm 2} with a good initialization.}\label{f2}
\end{figure}

\begin{figure}
\centerline{\includegraphics[width=20pc]{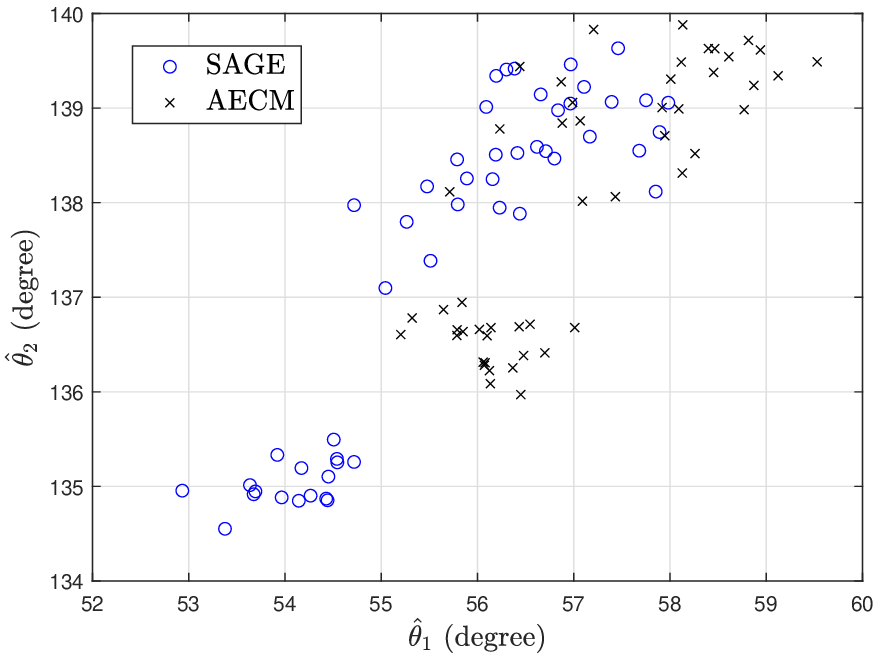}}
\caption{Scatter plot of $(\hat{\theta}_1,\hat{\theta}_2)$’s from both algorithms using \textbf{Algorithm 2} with a bad initialization.}\label{f3}
\end{figure}

Fig. \ref{f2} and Fig. \ref{f3} illustrate two scatter plots of $(\hat{\theta}_1,\hat{\theta}_2)$’s from both algorithms under $50$ independent realizations. Here, the same $\mathbf{Y}$ of each realization are processed by both algorithms. In Fig. \ref{f3}, we initialize $\hat{\theta}_1=45^{\circ}$ and $\hat{\theta}_2=120^{\circ}$, i.e., a bad initialization. It can be seen that all of the $(\hat{\theta}_1,\hat{\theta}_2)$’s in Fig. \ref{f2} center around the true value $(60^{\circ},100^{\circ})$ due to the good initialization. On the contrary, all of the $(\hat{\theta}_1,\hat{\theta}_2)$’s in Fig. \ref{f3} deviate from the true value $(60^{\circ},100^{\circ})$ due to the bad initialization. Thus, both algorithms can converge to the global optimum of problem \eqref{7} under a good initialization.

\subsection{Accuracy Comparison}

In this subsection, we compare and prove the accuracy of ML based DOA estimation. According to the above simulation results, we only use the AECM algorithm using \textbf{Algorithm \ref{alg:D}} to obtain the ML based $\hat{\theta}_m$'s, which requires the initialization to be good. In addition, we simulate two subspace based robust DOA estimation methods, namely, FLOM-MUSIC \cite{THLiu} and $\ell_p$-MUSIC \cite{Zeng}. Here, the root-MUSIC algorithm \cite{barabell,BDRao} is employed to obtain the $\hat{\theta}_m$'s after the signal or noise subspace estimate is obtained.

\begin{figure} 
\centerline{\includegraphics[width=20pc]{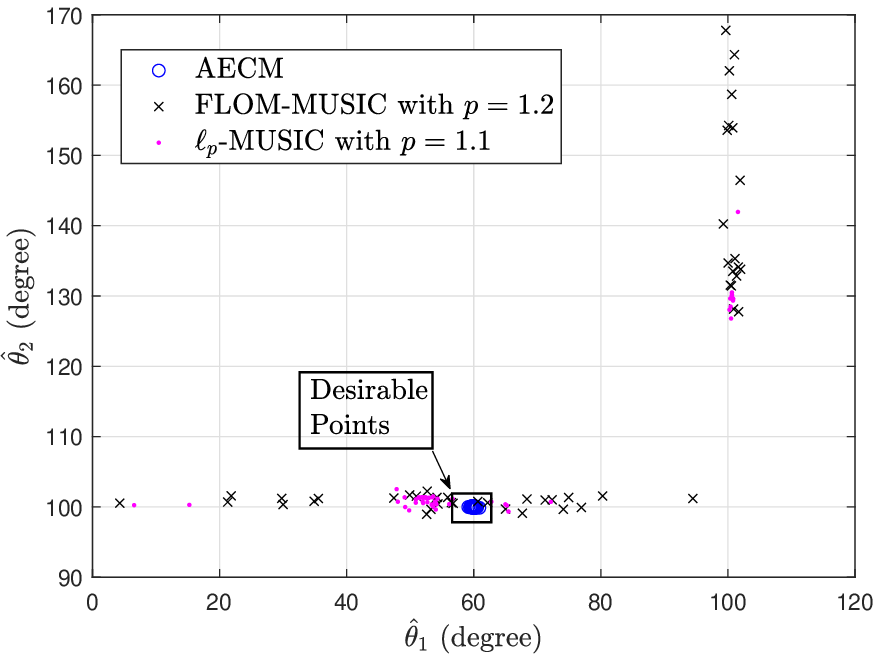}}
\caption{Scatter plot of $(\hat{\theta}_1,\hat{\theta}_2)$’s from different methods with coherent signals in Gaussian mixture noise.}\label{f4}
\end{figure}

\begin{figure} 
\centerline{\includegraphics[width=20pc]{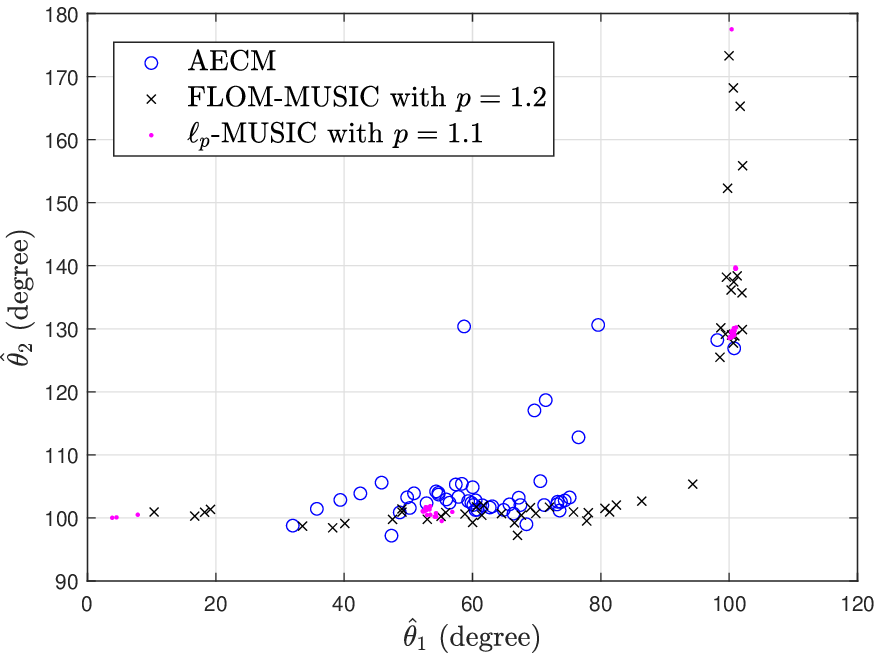}}
\caption{Scatter plot of $(\hat{\theta}_1,\hat{\theta}_2)$’s from different methods with coherent signals in $S\alpha S$ noise.}\label{f5}
\end{figure}

Fig. \ref{f4} and Fig. \ref{f5} illustrate two scatter plots of $(\hat{\theta}_1,\hat{\theta}_2)$’s from different methods under 50 independent realizations and the same $\mathbf{Y}$ of each realization is manipulated by these methods. We can see that in Fig. \ref{f4}, all of the $(\hat{\theta}_1,\hat{\theta}_2)$'s from AECM are desirable points while most of the $(\hat{\theta}_1,\hat{\theta}_2)$'s from FLOM-MUSIC and $\ell_p$-MUSIC are not desirable points. In Fig. \ref{f5}, we initialize $\hat{\sigma}_1=\hat{\sigma}_2=\sqrt{20}$ and adopt the complex $S\alpha S$ noise with characteristic exponent $\alpha=1.3$, scale parameter $\sigma=1$, skewness parameter $\beta=0$, and shift parameter $\mu=0$ \cite{THLiu}. We can see that although the noise is not Gaussian mixture noise, i.e., noise mismatch, many $(\hat{\theta}_1,\hat{\theta}_2)$'s from AECM slightly deviate from the true value $(60^{\circ},100^{\circ})$. Hence, the AECM algorithm, unlike existing subspace based robust DOA estimation methods, can still be used when signals are coherent.

\begin{figure} 
\centerline{\includegraphics[width=20pc]{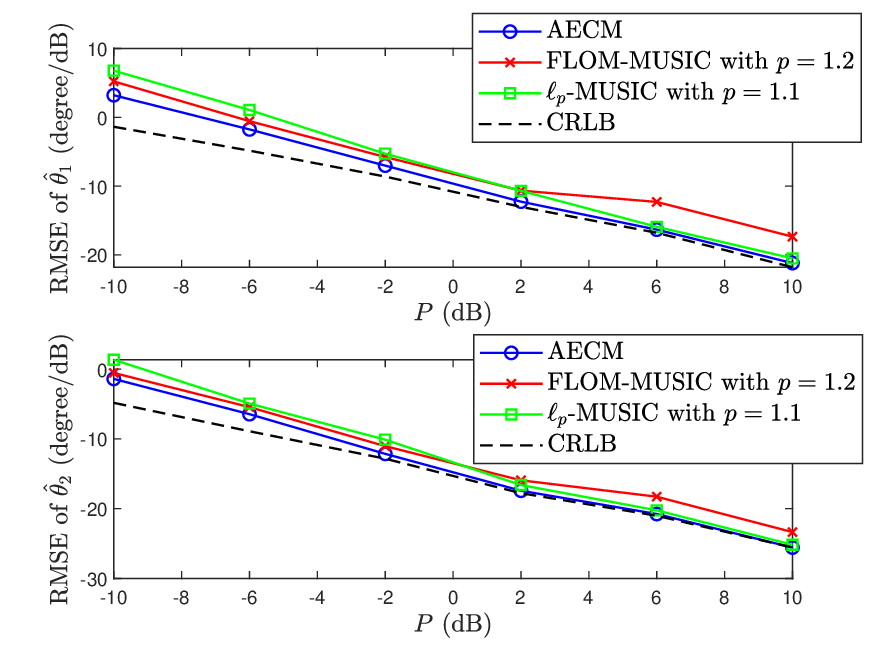}}
\caption{RMSE comparison of $(\hat{\theta}_1,\hat{\theta}_2)$ from different methods with noncoherent signals in Gaussian mixture noise.}\label{f6}
\end{figure}

\begin{figure}
\centerline{\includegraphics[width=20pc]{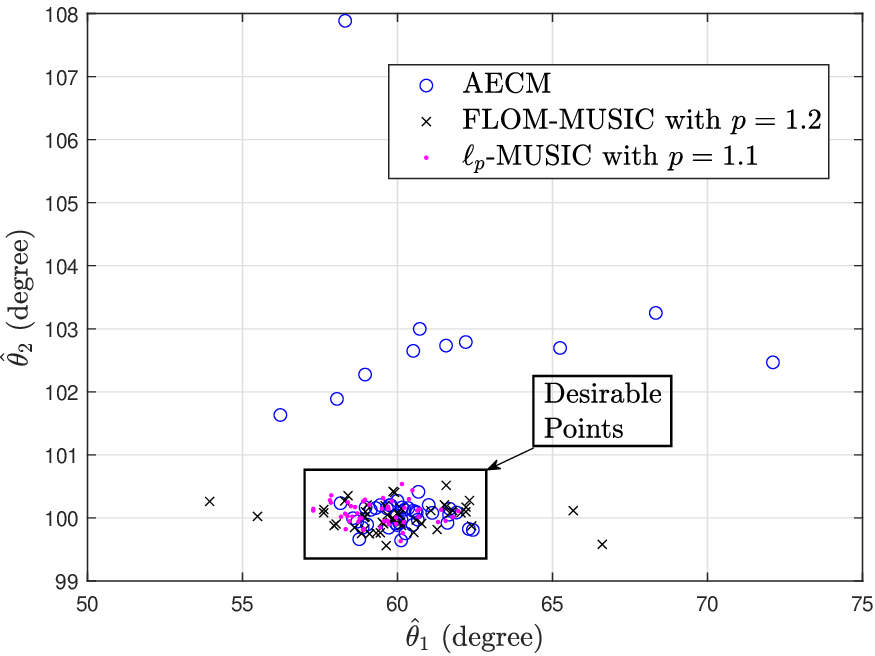}}
\caption{Scatter plot of $(\hat{\theta}_1,\hat{\theta}_2)$'s from different methods with noncoherent signals in $S\alpha S$ noise.}\label{f7}
\end{figure}

In Fig. \ref{f6} and Fig. \ref{f7}, we adopt noncoherent signals via $\mathbf{s}(t)\sim\mathcal{CN}\big(\mathbf{0},\mathrm{diag}\{P,10P\}\big)$ in order to let the subspace based robust DOA estimation methods work properly. Furthermore, each root mean squared error (RMSE) is calculated from $200$ independent realizations, i.e.,
\begin{eqnarray}
    \mathrm{RMSE}(\hat{\theta}_m)&=&\sqrt{\frac{1}{200}\sum_{i=1}^{200}\Big(\hat{\theta}_m^i-\theta_m\Big)^2}\nonumber\\
    &=&10\lg\Big[\frac{1}{200}\sum_{i=1}^{200}\Big(\hat{\theta}_m^i-\theta_m\Big)^2\Big]~(\mathrm{dB})
\end{eqnarray}
where $m=1,\dots,M$ and $\hat{\theta}_m^i$ is the $\hat{\theta}_m$ obtained from the $i$th independent realization. Fig. \ref{f6} shows RMSE results of $(\hat{\theta}_1,\hat{\theta}_2)$ from different methods with the Cramer–Rao lower bounds (CRLBs) \cite{Kozick}. As expected, we can see that the RMSE results of $(\hat{\theta}_1,\hat{\theta}_2)$ from AECM asymptotically achieve the CRLBs as $P$\footnote{Here, $P$ can be regarded as signal-to-noise ratio.} increases, and are smaller than those from FLOM-MUSIC and $\ell_p$-MUSIC. This is because ML based DOA estimation can offer the highest advantage in terms of accuracy without noise mismatch \cite{Krim}.

Fig. \ref{f7} illustrates a scatter plot of $(\hat{\theta}_1,\hat{\theta}_2)$’s from different methods under $50$ independent realizations and the same $\mathbf{Y}$ of each realization is manipulated by these methods. Here, we initialize $\hat{\sigma}_1=\sqrt{20}$ and $\hat{\sigma}_2=\sqrt{30}$, and adopt the complex $S\alpha S$ noise with characteristic exponent $\alpha=1.8$, scale parameter $\sigma=1$, skewness parameter $\beta=0$, and shift parameter $\mu=0$ \cite{THLiu}. In Fig. \ref{f7}, the numbers of desirable points from AECM, FLOM-MUSIC, and $\ell_p$-MUSIC are $39$, $46$, and $50$, respectively, which indicates that in the presence of $S\alpha S$ noise, i.e., noise mismatch, the $\hat{\theta}_m$'s based on problem \eqref{7} or ML are inferior to those from FLOM-MUSIC or $\ell_p$-MUSIC. This observation is consistent with the finding in \cite{Zeng}.

\begin{remark}
    We do not plot RMSE results of different methods in $S\alpha S$ noise since in the SAGE or AECM algorithm, updating the conditional probabilities of the $h_n(t)$'s may arise the meaningless situation of ``$0/0$'' due to exponential operation (i.e., $e^{-\vert x\vert^2}\to0$ when $\vert x\vert$ is large due to outliers) and finite precision, which causes algorithm breakdown. More importantly, $S\alpha S$ noise does not involve the parameters of $\boldsymbol{\lambda}$ and $\boldsymbol{\sigma}$, which makes the initialization of $\hat{\boldsymbol{\lambda}}$ and $\hat{\boldsymbol{\sigma}}$ in the SAGE or AECM algorithm unjudgeable. In other words, any initialization of $\hat{\boldsymbol{\lambda}}$ and $\hat{\boldsymbol{\sigma}}$ cannot guarantee that the SAGE or AECM algorithm converges to the global optimum of problem \eqref{7} when the noise is not Gaussian mixture noise. Moreover, unlike Fig. \ref{f6}, we cannot use the relation between RMSE results from the SAGE or AECM algorithm and the CRLBs to verify that the algorithm can converge to the global optimum of problem \eqref{7} because of noise mismatch. In conclusion, the SAGE or AECM algorithm is not recommended in the presence of $S\alpha S$ noise.
\end{remark}

\section{Conclusion}

The AECM algorithm has been applied and designed to efficiently solve the deterministic ML direction finding problem in Gaussian mixture noise, which utilizes multiple less informative versions of the complete data and the golden section search method to update DOA estimates at each iteration sequentially (one by one). Theoretical analysis has revealed that the AECM algorithm possesses nearly the same computational complexity of each iteration as the SAGE algorithm. However, simulation results have proven that the AECM algorithm exhibits faster stable convergence and is more attractive in computation. Importantly, the AECM algorithm, unlike existing subspace based robust DOA estimation methods, can still be used when signals are coherent.

\printbibliography 

@article{Godara,
  author        = {L. C. Godara},
  title         = {Application of antenna arrays to mobile communications. Part II: Beam-forming and direction-of-arrival considerations},
  journal       = "Proc. IEEE",
  volume        = "85",
  number        = "8",
  month         = {8},
  year          = "1997",
  pages         = "1195-1245"
}

@article{Miller,
  author        = {K. S. Miller},
  title         = {Complex Gaussian processes},
  journal       = "SIAM Rev.",
  volume        = "11",
  number        = "4",
  month         = {10},
  year          = "1969",
  pages         = "544-567"
}

@article{Krim,
  author        = {H. Krim and M. Viberg},
  title         = {Two decades of array signal processing research: The parametric approach},
  journal       = "IEEE Signal Process. Mag.",
  volume        = "13",
  number        = "4",
  month         = {7},
  year          = "1996",
  pages         = "67-94"
}

@article{Swami,
  author        = {A. Swami and B. M. Sadler},
  title         = {On some detection and estimation problems in heavy-tailed noise},
  journal       = "Signal Process.",
  volume        = "82",
  number        = "12",
  month         = {9},
  year          = "2002",
  pages         = "1829--1846"
}

@article{Tsakalides,
  author        = {P. Tsakalides and C. L. Nikias},
  title         = {The robust covariation-based MUSIC (ROC-MUSIC) algorithm for bearing estimation in impulsive noise environments},
  journal       = "IEEE Trans. Signal Process.",
  volume        = "44",
  number        = "7",
  month         = {7},
  year          = "1996",
  pages         = "1623--1633"
}

@inproceedings{barabell,
  author        = "A. J. Barabell",
  title         = "Improving the resolution performance of eigenstructure-based direction-finding algorithms",
  booktitle     = "IEEE International Conference on Acoustics, Speech, and Signal Processing",
  address       = "Boston, MA, USA",
  month         = 4,
  year          = "1983",
  pages         = "336-339"
}

@article{THLiu,
  author        = {T. H. Liu and J. M. Mendel},
  title         = {A subspace-based direction finding algorithm using fractional lower order statistics},
  journal       = "IEEE Trans. Signal Process.",
  volume        = "49",
  number        = "8",
  month         = {8},
  year          = "2001",
  pages         = "1605--1613"
}

@article{BDRao,
  author        = {B. D. Rao and K. V. S. Hari},
  title         = {Performance analysis of root-MUSIC},
  journal       = "IEEE Trans. Acoust., Speech, Signal Process.",
  volume        = "37",
  number        = "3",
  month         = {12},
  year          = "1989",
  pages         = "1939–1949"
}

@article{Visuri,
  author        = {S. Visuri and H. Oja and V. Koivunen},
  title         = {Subspace-based direction-of-arrival estimation using nonparametric statistics},
  journal       = "IEEE Trans. Signal Process.",
  volume        = "49",
  number        = "9",
  month         = {9},
  year          = "2001",
  pages         = "2060--2073"
}

@article{Zeng,
  author        = {W. Zeng and H. C. So and L. Huang},
  title         = {$\ell_p$-MUSIC: Robust direction-of-arrival estimator for impulsive noise environments},
  journal       = "IEEE Trans. Signal Process.",
  volume        = "61",
  number        = "17",
  month         = {9},
  year          = "2013",
  pages         = "4296--4308"
}

@article{Raj,
  author        = {P. Rajpurohit and P. Babu and P. Stocia},
  title         = {Robust direction-of-arrival estimation in the presence of outliers},
  journal       = "IEEE Trans. Aerosp. Electron. Syst.",
  volume        = "61",
  number        = "4",
  month         = {4},
  year          = "2025",
  pages         = "10921--10927"
}

@book{GJMcLachlan,
  author        = "G. J. McLachlan and D. Peel",
  title         = "Finite Mixture Models",
  publisher     = "Wiley-Interscience",
  year          = "2000"
}

@article{Redner,
  author        = {R. A. Redner and H. F. Walker},
  title         = {Mixture densities, maximum likelihood and the EM algorithm},
  journal       = "SIAM Rev.",
  volume        = "26",
  number        = "2",
  month         = {4},
  year          = "1984",
  pages         = "195--239"
}

@article{Dempster,
  author        = {A. P. Dempster and N. M. Laird and D. B. Rubin},
  title         = {Maximum likelihood from incomplete data via the EM algorithm},
  journal       = "J. R. Stat. Soc., Ser. B",
  volume        = "39",
  number        = "1",
  month         = {9},
  year          = "1977",
  pages         = "1--38"
}

@book{GMcLachlan,
  author        = "G. McLachlan and T. Krishnan",
  title         = "The EM Algorithm and Extensions",
  publisher     = "Wiley-Interscience",
  year          = "2008"
}

@article{Fessler,
  author        = {J. A. Fessler and A. O. Hero},
  title         = {Space-alternating generalized expectation-maximization algorithm},
  journal       = "IEEE Trans. Signal Process.",
  volume        = "42",
  number        = "10",
  month         = {10},
  year          = "1994",
  pages         = "2664--2677"
}

@article{Kozick,
  author        = {R. J. Kozick and B. M. Sadler},
  title         = { Maximum-likelihood array processing in non-Gaussian noise with Gaussian mixtures},
  journal       = "IEEE Trans. Signal Process.",
  volume        = "48",
  number        = "12",
  month         = {12},
  year          = "2000",
  pages         = "3520--3535"
}

@article{Song,
  author        = {J. Song and L. Cao and Z. Zhao and K. Yang and D. Wang and C. Fu},
  title         = {Adaptive gridless covariance-matching DOA estimation for coprime array under impulsive noise},
  journal       = "IEEE Trans. Instrum. Meas.",
  volume        = "75",
  month         = {1},
  year          = "2026",
  pages         = "1--17"
}

@ARTICLE{LiaoYP,
  author={Liao, Yanping and Qin, Chongxiang and Guo, Qiang},
  journal={IEEE Trans. Instrum. Meas.}, 
  title={GH-MUSIC: A Novel Accelerated DOA Estimation Method in Impulsive Noise Environments}, 
  year={2025},
  month={11},
  volume={75},
  pages={1-17}
  }

@article{Meng,
  author        = {X. Meng and D. V. Dyk},
  title         = {The EM algorithm---an old folk-song sung to a fast new tune},
  journal       = "J. R. Stat. Soc., Ser. B",
  volume        = "59",
  number        = "3",
  month         = {12},
  year          = "1997",
  pages         = "511--567"
}

@article{Ziskind,
  author        = {I. Ziskind and M. Wax},
  title         = {Maximum likelihood localization of multiple sources by alternating projection},
  journal       = "IEEE Trans. Acoust., Speech, Signal Process.",
  volume        = "36",
  number        = "10",
  month         = {10},
  year          = "1988",
  pages         = "1553--1560"
}

@book{Kay,
  author        = "S. M. Kay",
  title         = "Fundamentals of Statistical Signal Processing: Estimation Theory",
  publisher     = "Prentice Hall",
  year          = "1993"
}

@book{Samorodnitsky,
  author        = "G. Samorodnitsky",
  title         = "Stable Non-Gaussian Random Processes: Stochastic Models with Infinite Variance",
  publisher     = "Routledge",
  year          = "1994"
}

@article{Wu,
  author        = {C. F. J. Wu},
  title         = {On the convergence properties of the EM algorithm},
  journal       = "Ann. Statist.",
  volume        = "11",
  number        = "1",
  month         = {3},
  year          = "1983",
  pages         = "95--103"
}

@article{Shao,
  author        = {M. Shao and C. L. Nikias},
  title         = {Signal processing with fractional lower order moments: Stable processes and their applications},
  journal       = "Proc. IEEE",
  volume        = "81",
  number        = "7",
  month         = {7},
  year          = "1993",
  pages         = " 986–1010"
}

@BOOK{Nikias,
        title = {Signal Processing with Alpha-Stable Distributions and Applications},
	publisher = {Wiley-Interscience},
	author = {L. Nikias and M. Shao},
	year = {1995}
}

@BOOK{Boyd,
        title = {Convex Optimization},
	publisher = {Cambridge University Press},
	author = {Boyd S. and Vandenberghe L.},
	year = {2004}
}

@ARTICLE{Gong,
  author={Gong, Ming-yan and Lyu, Bin},
  journal={IEEE Trans. Veh. Technol.}, 
  title={Alternating Maximization and the EM Algorithm in Maximum-Likelihood Direction Finding}, 
  year={2021},
  volume={70},
  month={8},
  number={10},
  pages={9634-9645},
}

\end{document}